\documentstyle[11pt,stestyle]{article}

\title{An operator-based exact treatment of open quantum systems}
\author{S.Nicolosi\thanks{I thank Prof.N.Messina and Dr.A.Napoli to introduce me on the subject and to give me the freedom to follow my personal scientific interest.} \\{\footnotesize\it stefania.nicolosi@virgilio.it\\INFM, MIUR and Dipartimento di Scienze Fisiche ed
Astronomiche, via Archirafi 36, 90123 Palermo, Italy, Tel: +39 091
6234248, Fax: +39 091 6234281 } }

\begin{document}

\maketitle
\begin{abstract}
{\it Quantum mechanics must be regarded as open systems. On one
hand, this is due to the fact that, like in classical physics, any
realistic system is subjected to a coupling to an uncontrollable
environment which influences it in a non-negligible way. The
theory of open quantum systems thus play a major role in many
applications of quantum physics since perfect isolation of quantum
system is not possible and since a complete microscopic
description or control of the environment degrees of freedom is
not feasible or only partially so} \cite{F.Petruccione}. Practical
considerations therefore force one to seek for a simpler,
effectively probabilistic description in terms of an open system.
There is a close physical and mathematical connection between the
evolution of an open system, the state changes induced by quantum
measurements, and the classical notion of a stochastic process.
The paper {\it provides} a bibliographic review of this
interrelations, it {\it shows} the mathematical equivalence
between {\it markovian master equation} and generalized {\it
piecewise deterministic processes} \cite{F.Petruccione} and it
{\it introduces} the {\it open system in an open observed
environment} model.
\end{abstract}

\section{Introduction}
The dynamics of open quantum systems plays a central role in a
wide class of physical systems. Usually, the dynamics of an open
system is described in terms of the reduced density matrix $\rho_S
(t)$ which is defined by the trace over the environment degrees of
freedom \cite{F.Petruccione,Gardiner,Louiselle}. On the ground of
the weak-coupling assumption and the Rotating Wave Approximation
the dynamics may be formulated in terms of a quantum dynamical
semigroups which yields a Markovian Master Equation \cite{Alicki}.
However, the dynamical equation, thus obtained, is very often
untractable. This fact has encouraged some physicists
\cite{F.Petruccione,Carmichael,Molmer1,Molmer2} to look for
alternative ways to describe open systems. Instead of representing
the dynamics of an open system by a quantum Master Equation for
its density matrix, it is formulated in terms of a stochastic
process for the open system's wave function. The stochastic
representation of quantum Markov processes already appeared in a
fundamental paper by Davies \cite{Davies} and was applied
\cite{Davies2} to derive the {\it photocounting formula}. While
the theory was originally formulated in terms of a stochastic
process for the reduced density matrix, in the last decade it has
been proposed \cite{F.Petruccione,Carmichael,Molmer1,Molmer2} as a
stochastic evolution of the state vector in the reduced Hilbert
space (for a review see \cite{Plenio-Knight}). At the same time
Carmichael has developed the idea of an {\it unravelling} of the
master equation in terms of an ensemble of {\it quantum
trajectories}. His theory is applicable only to a particular class
of quantum systems (the photoemissive sources) and it induces to
think that this treatment is equivalent to the Master Equation
approach. In the last few years F.Petruccione and H.P.Breuer
\cite{F.Petruccione} generalize and provide a mathematical
formulation of Carmichael's idea of {\it quantum trajectories}
\cite{Carmichael} and of the {\it Monte Carlo wave function
method} \cite{Molmer1}. In this contest their main result is to
demonstrate that the dynamics given by the most general Master
Equation in Lindblad form can be represented as a {\it piecewise
deterministic process} (PDP) \cite{F.Petruccione} $\psi(t)$ in the
Hilbert space of the open system. The physical basis to achieve
their aim is provided by continuous measurement theory
\cite{F.Petruccione}.

The link \cite{diagramma} between the first and second way of
describing open quantum systems is, however, only one way in the
sense I am going to explain. The cited paper contains a flow
diagram that is a clear and straightforward picture to review {\it
Concept and methods in the theory of open quantum systems}, as the
same authors title their work. The diagram shows us that PDPs,
describing a {\it selective} level of measurement, implies the
{\it non-selective one}, appearing in the form of a Lindblad
markovian master equation, while the opposite has not been
demonstrated. My aim is to put an arrow in the opposite direction
in order to demonstrate the equivalence between the two approaches
to open systems dynamics, at least under Born-Markov
approximation. To achieve my goal I start from microscopic models
and, exploiting the same approximation leading to the most general
Master Equation (not in Lindblad form \footnote{Every Markovian
Master Equation can be put in Lindblad form and this, in general,
introduce simplification in the further calculations, but because
of the difficulty to recast the equation in this form the results
obtained are in general merely formal. For this reason I prefer to
work with non-diagonal Markovian Master Equation, by virtue of
which my solution is applicable to the more of the known
open-system theoretical models -from matter-radiation interaction
model to the spin-boson model-, more easily than other
approaches.}), I solve this last exactly (at a bath temperature
$T=0$) obtaining an operatorial expression for $\rho_S(t)$.
Moreover, in the contest of optical quantum system I derive an
expression for $\rho_S(t)$ that is in accordance with the
Carmichael's one, but, differently from that, the mine is
applicable also when the Master equation is not in the Lindblad
form in which cases Carmichael's solution is very often merely
formal, as the same author underline in recent papers
\cite{Carm2000,Carm2003}. My mathematical tool, here called {\it
NuD Theorem}, is applicable to a wide class of systems, provided
that they satisfy the hypothesis necessary to make it working.
Optical physical systems well satisfy the conditions of NuD
theorem's validity and in this contest I have studied, as example
of monopartite system, a single mode cavity, and as example of
multipartite systems, two two-level {\it dipole-dipole}
interacting atoms and $N$ two-level not-directly-interacting atoms
placed in fixed \textit{arbitrary} point inside a loss cavity
\cite{Nicolosi,NicolosiPRL}. In the investigation of the
monopartite systems the novelty is constituted by the operatorial
way to approach to the master equations of the systems, being
already known in literature
\cite{F.Petruccione,Gardiner,Louiselle,Barnett2002} their
dynamical properties. I  reproduce for example the {\it
photocounting formula} in order to appreciate the easyness of
application of my method. Moreover I choose to analyze monopartite
systems at a temperature $T=0$ to highlight that, differently to
multipartite ones, spontaneous emission provokes decoherence
phenomena that, inevitably, guides the first kind of systems to
their ground states. On the contrary, multipartite systems can
exhibit collective properties induced by the common reservoir
\cite{Nicolosi,NicolosiPRL,Leonardi,subradiant
state,Kozierowski,Shore,Kudryavtsev,Phoenix}. This general feature
already appeared in some fundamental papers \cite{Dicke,Leonardi}
in which the interaction between atomic dipoles, induced by
electromagnetic field, could cause the decay of the multiatom
system with two significantly different spontaneous emission
rates, one enhanced and the other reduced.

In two recent papers, as an example of multipartite system, I have
investigated the dynamics of a couple of spontaneously emitting
two-level atoms, taking into account from the very beginning their
dipole-dipole interaction and and $N$ two-level
not-directly-interacting atoms placed in fixed \textit{arbitrary}
point inside a loss cavity \cite{Nicolosi,NicolosiPRL}. The
result, not trivially expected, is that in such a condition the
matter subsystem, because of the cooperation induced by energy
loss mechanism, may be conditionally guided toward a stationary
robust entangled state. The renewed interest toward entanglement
concept reflects the consolidated belief that unfactorizable
states of multipartite system provide an unreplaceable applicative
resource, for example, in the quantum computing  research area
\cite{Nielsen}. However, the realization of quantum computation
protocols suffers of the difficulty of isolating a quantum
mechanical system from its environment. In this sense the cited
work are also aimed at proposing theoretical scheme to bypass
decoherence manifestations, so taking its place among intense
theoretical and experimental research of the last few years
\cite{Nicolosi,NicolosiPRL,Cinesi,Cinesi2,Cirac,Shneider,Yang,Hagley,Foldi}
\footnote{Citations \cite{Nicolosi, NicolosiPRL} represent my
first results about noise-induced entanglement. Even if they are
obtained solving the coupled differential equation system of the
block vector components describing the open system evolution, they
can be considered the first NuD theorem application because the
way in which I obtain the solution is always the same and it find
its generalization in the NuD theorem.}.

The paper is structured as follow: in section II I report the
principal step and approximation leading to the microscopic
derivation of the Markovian Master Equation, putting in evidence
some peculiar properties of it useful in order to demonstrate the
{\it NuD Theorem N}. In section III I solve the markovian master
equation when $T=0$. In section IV I review the applications to
old exemplary problem and to new, previous unresolved, problem. In
section V I try to justify the obtained dynamical behaviour in
terms of continuous measurement theory.

\section{Quantum Markovian Master Equation}
It is well know that under the Rotating Wave and the Born-Markov
approximations the master equation describing the reduced
dynamical behavior of a generic quantum system linearly coupled to
an environment can be put in the form \cite{F.Petruccione}
\begin{equation}\label{ME}
\dot{\rho_S} (t)=-i[H_S+H_{LS},\rho_S (t)]+D(\rho_S (t)),
\end{equation}
where $H_S$ is the hamiltonian describing the free evolution of
the isolated system,
\begin{eqnarray}\label{dissME}
\nonumber D(\rho_S (t))&&=\sum_{\omega } \sum_{\alpha ,\beta}
 \gamma_{\alpha ,\beta} (\omega)
(A_{\beta} (\omega ) \rho_S (t) A^{\dag}_{\alpha} (\omega
)\\&&-\frac{1}{2}\{A^{\dag}_{\alpha} (\omega ) A_{\beta} (\omega
), \rho_S (t)\}),
\end{eqnarray}

\begin{equation}\label{lambshME}
 H_{LS} =\sum_{\omega}\sum_{\alpha,\beta}S_{\alpha ,\beta}
 (\omega)A^\dag_{\alpha}(\omega)A_{\beta}(\omega),
\end{equation}

\begin{equation}
 S_{\alpha ,\beta} (\omega)=\frac{1}{2i}(\Gamma_{\alpha ,\beta}
 (\omega)-\Gamma^\ast_{\beta ,\alpha}(\omega))
\end{equation}
and
\begin{equation}
 \gamma_{\alpha ,\beta} (\omega)=\Gamma_{\alpha ,\beta}
 (\omega)+\Gamma^\ast_{\beta ,\alpha}(\omega),
\end{equation}

$\Gamma_{\alpha ,\beta}(\omega)$ being the one-sided Fourier
transforms of the reservoir correlation functions. Finally we
recall that the operators $A_{\alpha}(\omega)$ and
$A_{\alpha}^{\dag}(\omega)$, we are going to define and whose
properties we are going to explore, act only in the Hilbert space
of the system.

Eq.\ (\ref{ME}) has been derived under the hypothesis that the
interaction hamiltonian between the system and the reservoir, in
the Schr\"{o}dinger picture, is given by \cite{F.Petruccione}

\begin{equation}\label{HI}
H_I=\sum_{\alpha} A^{\alpha}\otimes B^{\alpha},
\end{equation}
that is the most general form of the interaction.

In the above expression $A^{\alpha}=(A^{\alpha})^\dag$ and
$B^{\alpha}=(B^{\alpha})^\dag$ are operators acting respectively
on the Hilbert space of the system  and  of the reservoir. The
eq.\ (\ref{HI}) can be written in a slightly different form if one
decomposes the interaction hamiltonian into eigenoperators of the
system and reservoir free hamiltonian.

\begin{definition}{Definition}
Supposing the spectrums of $H_S$ and $H_B$ to be discrete
(generalization to the continuous case is trivial) let us denote
the eigenvalue of $H_S$ ($H_B$) by $\varepsilon$ ($\eta$) and the
projection operator onto the eigenspace belonging to the
eigenvalue $\varepsilon$ ($\eta$) by $\Pi (\varepsilon)$ ($\Pi
(\eta)$). Then we can define the operators:
\begin{equation}\label{AdiOmega}
A_{\alpha} (\omega)\equiv\sum_{\varepsilon^{'}
-\varepsilon=\omega} \Pi (\varepsilon) A_{\alpha}\Pi
(\varepsilon^{'}),
\end{equation}
\begin{equation}\label{AdiOmega}
B_{\alpha} (\omega)\equiv\sum_{\eta^{'} -\eta=\omega} \Pi (\eta)
B_{\alpha}\Pi (\eta^{'}).
\end{equation}
\end{definition}

From the above definition we immediately deduce the following
relations

\begin{equation}\label{com1}
[H_S, A_{\alpha} (\omega)]=-\omega A_{\alpha} (\omega),\;\;\;
[H_B, B_{\alpha} (\omega)]=-\omega B_{\alpha} (\omega),
\end{equation}

\begin{equation}\label{com2}
[H_S, A^{\dag}_{\alpha} (\omega)]=+\omega A^{\dag}_{\alpha}
(\omega)\;\;\; and \;\;\; [H_B, B^{\dag}_{\alpha}
(\omega)]=+\omega B^{\dag}_{\alpha} (\omega).
\end{equation}
An immediate consequence is that the operators $A^{\dag}_{\alpha}
(\omega>0)$ e $A_{\alpha} (\omega>0)$ raise and lower the energy
of the system $S$ by the amount $\hbar \omega$ respectively and
that the corresponding interaction picture operators take the form

\begin{equation}
e^{iH_{S} t}A_{\alpha} (\omega) e^{-iH_{S} t}=e^{-i\omega t}
A_{\alpha} (\omega), \;\;\; e^{iH_{B} t}B_{\alpha} (\omega)
e^{-iH_{B} t}=e^{-i\omega t} B_{\alpha} (\omega),
\end{equation}

\begin{equation}
e^{iH_{S} t}A^{\dag}_{\alpha} (\omega) e^{-iH_{S} t}=e^{+i\omega
t} A^{\dag}_{\alpha} (\omega) \;\;\; and \;\;\; e^{iH_{B}
t}B^{\dag}_{\alpha} (\omega) B^{-iH_{B} t}=e^{+i\omega t}
B^{\dag}_{\alpha} (\omega).
\end{equation}

Finally we note that
\begin{equation}\label{AconAlfadiOmega}
A^{\dag}_{\alpha} (\omega)=A_{\alpha} (-\omega) \;\;\; and \;\;\;
B^{\dag}_{\alpha} (\omega)=B_{\alpha} (-\omega).
\end{equation}
Summing eq.\ (\ref{AconAlfadiOmega}) over all anergy differences
and employing the completeness relation we get

\begin{equation}
\sum_{\omega}A^{\dag}_{\alpha} (\omega)=\sum_{\omega}A_{\alpha}
(-\omega)=A_\alpha  \;\;\; and
\;\;\;\sum_{\omega}B^{\dag}_{\alpha}
(\omega)=\sum_{\omega}B_{\alpha} (-\omega)=B_\alpha
\end{equation}
The above positions enable us to cast the interaction hamiltonian
into the following form
\begin{equation}
H_I=\sum_{\alpha ,\omega,\omega '} A_{\alpha} (\omega) \otimes
B_{\alpha}(\omega ')=\sum_{\alpha ,\omega,\omega '}
A^{\dag}_{\alpha} (\omega) \otimes B^{\dag}_{\alpha}(\omega ').
\end{equation}
The reason for introducing the eigenoperator decomposition, by
virtue of which  the interaction hamiltonian in the interaction
picture can now be written as
\begin{equation}\label{intgen}
H_I(t)=\sum_{\alpha ,\omega,\omega '}e^{-i(\omega+\omega ')t}
A_{\alpha} (\omega) \otimes B_{\alpha}(\omega '),
\end{equation}
is that exploiting the rotating wave approximation, whose
microscopic effect is to drop the terms for which $\omega\neq
-\omega'$, is equivalent to the schrodinger picture interaction
hamiltonian:
\begin{equation}
H_I=\sum_{\alpha ,\omega} A_{\alpha} (\omega) \otimes
B_{\alpha}(-\omega)=\sum_{\alpha ,\omega} A_{\alpha} (\omega)
\otimes B^{\dag}_{\alpha}(\omega).
\end{equation}

\begin{theorem}{Lemma} \label{TH1}
The Rotating Wave Approximation imply the conservation of the free
energy of the global system, that is
\begin{equation}
 [H_S+H_B,H]=0
\end{equation}

\end{theorem}

\subsection{Proof}
The necessary condition involved in the previous proposition is
equivalent to the equation $[H_S+H_B,H_I]=0$ we are going to
demonstrate.

\begin{eqnarray}\label{cons}
[H_S+H_B,H]&=&[H_S+H_B,H_I]=[H_S,H_I]+[H_B,H_I]\\\nonumber &
=&\sum_{\alpha ,\omega } [H_S,A_{\alpha} (\omega)] \otimes
B_{\alpha}^{\dag}(\omega) +\sum_{\alpha ,\omega } A_{\alpha}
(\omega) \otimes [H_B,B_{\alpha}^{\dag}(\omega)]\\\nonumber
&=&-\sum_{\alpha ,\omega}\omega A_{\alpha} (\omega) \otimes
B_{\alpha}(-\omega)+\sum_{\alpha ,\omega}\omega A_{\alpha}
(\omega) \otimes B_{\alpha}(-\omega)=0.
\end{eqnarray}

where we have made use of eq. (\ref{com1},\ref{com2})
\rule{5pt}{5pt}

\begin{theorem}{Lemma} \label{TH2}
The detailed balance condition in the thermodynamic limit imply
\cite{Alicki}
\begin{equation}
 \gamma_{\alpha \beta}(\omega)=e^{-\beta \omega} \gamma_{\alpha
 \beta}(-\omega)
\end{equation}
\end{theorem}
where $\beta=(k_B T)^{-1}$ \rule{5pt}{5pt}
\begin{theorem}{Corollary} \label{TH3}
Let us suppose the temperature of the thermal reservoir to be the
absolute zero, on the ground of Lemma 2 immediately we see that
\begin{equation}
 \gamma_{\alpha \beta}(\omega<0)=0\;\;\;\rule{5pt}{5pt}
\end{equation}
\end{theorem}

Let us now  cast eq.\ (\ref{ME}) in a slightly different form
splitting the sum over the frequency, appearing in eq.\
(\ref{dissME}), in a sum over the positive frequencies and a sum
over the negative ones so to obtain

\begin{eqnarray}\label{Diss}
&&\nonumber D(\rho_S (t))\\ \nonumber &=&\sum_{\omega>0 ,\alpha
,\beta} \gamma^{\alpha ,\beta} (\omega) (A^{\beta} (\omega) \rho_S
A^{\alpha \dag} (\omega)\\\nonumber &-&\frac{1}{2}\{A^{\beta \dag}
(\omega) A^{\alpha} (\omega),\rho_S \})\\\nonumber &+&\sum_{\omega
>0 ,\alpha ,\beta} \gamma^{\alpha ,\beta} (-\omega)
(A^{\alpha \dag} (\omega) \rho_S A^{\beta} (\omega)
\\&-&\frac{1}{2}\{A^{\alpha} (\omega) A^{\beta \dag} (\omega),\rho_S
\}),
\end{eqnarray}
where we again make use of eq.\ (\ref{AconAlfadiOmega}). In the
above expression we can recognize the first term as responsible of
spontaneous and stimulated emission processes, while the second
one takes into account stimulated absorption, as imposed by the
lowering and raising properties of $A^{\alpha}(\omega)$. Therefore
if the reservoir is a thermal bath at $T=0$ the corollary 4 tell
us that the correct dissipator of the Master Equation can be
obtained by suppressing the stimulated absorption processes in
eq.\ (\ref{Diss}).

\section{NuD Theorem}
We are now able to solve the markovian master equation when the
reservoir is in a thermal equilibrium state characterized by
$T=0$. We will solve a Cauchy problem assuming the factorized
initial condition to be an eigenoperator of the free energy
$H_S+H_B$. This hypothesis doesn't condition the generality of the
found solution being able to extend itself to an arbitrary initial
condition because of the linearity of the markovian master
equation \footnote{ It is out of relevance to consider initial
condition having non-zero coherence between the environment and
the system because it is not possible to resolve them in the
reduced dynamics obtained tracing on the environment degrees of
freedom.}.

\begin{theorem}{NuD theorem} \label{TH4}
If eq. (\ref{ME}) is the markovian master equation describing the
dynamical evolution of a open quantum system S, coupled to an
environment B, assumed to be in the detailed-balance thermal
equilibrium state characterized by a temperature T=0, and if the
global system is initially prepared in a state $\rho(0)=\rho_B(0)
\rho_S(0)$ so that $(H_S+H_B)\rho(0)(H_S+H_B)=E_L^2 \rho(0)$,
where $E_L=E_S+E_B$ is the free energy of the global system then
$\rho_S(t)$ is in the form of a Piecewise Deterministic Process
\cite{F.Petruccione}, that is a process obtained combining a
deterministic time-evolution with a jump process.
\end{theorem}

\subsection{Proof}
The weak-coupling assumption is equivalent to $\rho(t)=\rho_B(0)
\rho_S(t)$. The above equation can be used to derive the reduced
density matrix $\rho(t)$ tracing over the environment degree of
freedom. Let us choose a factorized base $B$ in the tensor product
Hilbert space made of eigenvectors of $H_S$ and $H_B$
\begin{eqnarray}
B=\{|E_B,\lambda_{E_B}>|E_S,\lambda_{E_S}>\},
\end{eqnarray}
where $\{E_B\}$ and $\{E_S\}$ define respectively the spectra of
$H_B$ and $H_S$ and $\{\lambda_{E_B}\}$ and $\{\lambda_{E_S}\}$
their relative degenerations. Let us remember that we have made
the semplificative hypotheses of discreteness of $\{E_B\}$ and
$\{E_S\}$. In addition we assume, also for easyness, that
$\{E_S\}$ is bounded from below and made of isolated points. On
the ground of these chioses the total density matrix can be
written as
\begin{eqnarray}
\rho(t)=\sum_{E_B,\lambda_{E_B},E_S,\lambda_{E_S},E'_{B},\lambda'_{E_B},E'_S,\lambda'_{E_S}}
\rho(E_B,\lambda_{E_B},E_S,\lambda_{E_S},E'_{B},\lambda'_{E_B},E'_S,\lambda'_{E_S},t)\\\nonumber
|E_B,\lambda_{E_B}>|E_S,\lambda_{E_S}><E'_B,\lambda'_{E_B}|<E'_S,\lambda'_{E_S}|.
\end{eqnarray}
The Lemma \ref{TH1} imposes a strong selection rule on the indices
of the summation, that is :
\begin{eqnarray}\label{selrule}
 E_S+E_B=E'_S+E'_B
\end{eqnarray}
by virtue of which the trace over the degrees of freedom of the
environment, that can be written as
\begin{eqnarray}
\rho_S(t)=\sum_{E_S,\lambda_{E_S},E'_S,\lambda'_{E_S}}
\rho(E_S,\lambda_{E_S},E'_S,\lambda'_{E_S},t)
|E_S,\lambda_{E_S}><E'_S,\lambda'_{E_S}|,
\end{eqnarray}
is immediately obtained:
\begin{eqnarray}
\rho_S(t)=\sum_{E_S}(\sum_{\lambda_{E_S},\lambda'_{E_S}}
\rho(E_S,\lambda_{E_S},\lambda'_{E_S},t)
|E_S,\lambda_{E_S}><E_S,\lambda'_{E_S}|)=\sum_{E_S}\rho_{E_S}(t).
\end{eqnarray}

The ensemble $\{E_S\}$ can be put in biunivocal correspondence
with the natural ensemble so that
\begin{eqnarray}\label{somma}
\rho_S(t)=\sum_{i=0}^N\rho_{i}(t),
\end{eqnarray}
where $N$ is the natural index corresponding to the maximum of
$\{E_S\}$. The operators $\rho_{i}(t)$ satisfy a big orthogonality
condition that is
\begin{eqnarray}
\rho_i(t)\cdot\rho_j(t)=\rho_i^2\delta_{ij}.
\end{eqnarray}
The last two equations demonstrate that the evolution is a
piecewise process (PP) or, equivalently, a statistical mixture of
alternative generalized trajectories. These last are {\it
generalized} respect to F.Petruccione and H.J.Carmochael approach,
which leads to $\rho_S(t)=\sum_i |\psi_i><\psi_i|$. The last
expansion, in terms of proper trajectories, is obtainable from the
mine if and only if we are able to dygonalize the spectral
correlation tensor, that is known to be always possible, but
nobody is able to do it, with exception of few highly symmetrical
systems. In order to demonstrate the formal equivalence between
the two approaches we have to demonstrate that the alternative
processes are deterministic (PDP) or, equivalently, that every of
them is representable in the form of an evolutionary equation.

On the ground of Lemma \ref{TH2} and its corollary the markovian
master equation at $T=0$ can be written as
\begin{equation}\label{ME0}
\dot{\rho_S} (t)=-i[H_S+H_{LS},\rho_S (t)]+D(\rho_S (t)),
\end{equation}

\begin{eqnarray}\label{dissME0}
\nonumber D(\rho_S (t))&&=\sum_{\omega>0 } \sum_{\alpha ,\beta}
 \gamma_{\alpha ,\beta} (\omega)
(A_{\beta} (\omega ) \rho_S (t) A^{\dag}_{\alpha} (\omega
)\\&&-\frac{1}{2}\{A^{\dag}_{\alpha} (\omega ) A_{\beta} (\omega
), \rho_S (t)\}).
\end{eqnarray}
Let us now substitute eq. (\ref{somma}) into eq. (\ref{ME0}), so
obtaining
\begin{eqnarray}\label{sist}
&&\sum_{i=0}^{N} \dot{\rho_i}
(t)=-\frac{i}{\hbar}\sum_{i=0}^{N}[H_0,\rho_i (t)]+\sum_{i,\omega
>0 ,\alpha ,\beta} \gamma^{\alpha ,\beta} (\omega)(A^{\beta}
(\omega) \rho_i A^{\alpha \dag} (\omega) \\
\nonumber &-&\frac{1}{2}\{A^{\beta \dag} (\omega) A^{\alpha}
(\omega),\rho_i \}).
\end{eqnarray}
$N$ being the natural index corresponding to the initial
eigenvalue of $H_S$ \footnote{At T=0 only spontaneous emission
processes are involved in the dynamics, so that the initial
eigenvalue of $H_S$ is the maximum permitted eigenvalue.}. Let us
observe that
\begin{equation}
\sum_{\omega
>0 ,\alpha ,\beta} \gamma^{\alpha ,\beta} (\omega)A^{\beta}
(\omega) \rho_0 A^{\alpha \dag} (\omega)=0
\end{equation}
so that in the second summation the index $i$ starts from $1$ and
then the eq. (\ref{sist}) can be written as

\begin{eqnarray}
\nonumber &\sum_{i=0}^{N-1}& [\dot{\rho_i}
(t)+\frac{i}{\hbar}[H_0,\rho_i (t)]\\ \nonumber &+&\sum_{\omega
>0 ,\alpha ,\beta} \gamma^{\alpha ,\beta} (\omega)(A^{\beta}
(\omega) \rho_{i+1} A^{\alpha \dag} (\omega)
+\frac{1}{2}\{A^{\beta \dag} (\omega) A^{\alpha} (\omega),\rho_i
\})]\\  &+& [\dot{\rho_N} (t)+\frac{i}{\hbar}[H_0,\rho_N
(t)]+\sum_{\omega
>0 ,\alpha ,\beta} \gamma^{\alpha ,\beta} \frac{1}{2}\{A^{\beta \dag} (\omega) A^{\alpha} (\omega),\rho_i
\}]=0.
\end{eqnarray}

Since the big orthogonality of $\{\rho_i\}$ the addenda in the
above equation act on disjointed subspaces of the Hilbert space of
$H_S$. This property implies that the above equation is verified
if and only if the below system of differential coupled equations
holds

\begin{eqnarray}
 &&\dot{\rho_i}
(t)=-\frac{i}{\hbar}[H_0,\rho_i (t)]\\ \nonumber &+&\sum_{\omega>0
,\alpha ,\beta} \gamma^{\alpha ,\beta} (\omega) (A^{\beta}
(\omega) \rho_{i+1} A^{\alpha \dag} (\omega)
-\frac{1}{2}\{A^{\beta \dag} (\omega) A^{\alpha} (\omega),\rho_i
\}).
\end{eqnarray}

It is immediate verify by substitution that

\begin{equation}
\rho_i (t)=U(t)f_i (t) U^{\dag} (t),
\end{equation}
where, in particular,
\begin{equation}
f_{N} (t)=\rho_{N} (0)
\end{equation}
and $U(t)=e^{-\frac{i}{\hbar}Bt}$, $U^\dag
(t)=e^{\frac{i}{\hbar}B^\dag t}$, $B$ being
\begin{equation}
 B=H_0-\frac{i}{2\hbar}\sum_{\omega>0 ,\alpha ,\beta}
\gamma^{\alpha \beta} (\omega)A^{\beta
\dag}(\omega)A^{\alpha}(\omega)\equiv H_0-\frac{i}{2\hbar}H',
\end{equation}
with $H'$ hermitian. Finally,
\begin{eqnarray}\label{brutta}
&f_{N-j}& (t)=\sum_{\omega ',\alpha ',\beta '}\sum_{\omega
",\alpha ",\beta "}...\sum_{\omega^j ,\alpha^j
,\beta^j}\gamma^{\alpha \beta} (\omega)\gamma^{\alpha ' \beta '}
(\omega ')\gamma^{\alpha "\beta "} (\omega ")...\gamma^{\alpha^j
\beta^j} (\omega^j) \\ \nonumber &\times & \int_0^t
\int_0^{t'}\int_0^{t"}...\int_0^{t^j}dt'dt"...dt^j
U^{-1}(t')A^{\beta '}(\omega ')U(t')U^{-1}(t")A^{\beta "}(\omega
")U(t")...\\ \nonumber &.&
U^{-1}(t^j)A^{\beta^j}(\omega^j)U(t^j)f_{N} (t^j)
U^{\dag}(t^j)A^{\alpha^j \dag}(\omega^j)U^{\dag -1}(t")... \\
\nonumber &.& U^{\dag}(t")A^{\alpha " \dag}(\omega ")U^{\dag
-1}(t")U^{\dag}(t')A^{\alpha ' \dag}(\omega ')U^{\dag -1}(t'), \;
\; \; \;  j=1,...,N \;\;\; \rule{5pt}{5pt}
\end{eqnarray}

This concludes the proof and, in addition, ensures that the
dynamical processes, whose statistical mixture gives the open
system stocastic evolution, are deterministic. This demonstrates
that the evolution is representable as a Piecewise Deterministic
Process (PDP) \cite{F.Petruccione}. The found solution generalizes
the PDPs introduced by H.J.Carmichael and formalized by
F.Petruccione and H.P.Breuer. Actually, it is applicable also when
the Markovian Master Equation isn't in the Lindblad form. This, as
already highlighted, in general, introduces simplification in the
further calculations, but because of the difficulty to recast the
equation in this form the results obtained are in general merely
formal. Tough the eq. (\ref{brutta}) seems complicated to use it
is a powerful predictive tool. I have tested it in a lot of
contests. Two of my results are already published, others will be
object of future papers. All of them contain the same
(predictable?) result: multipartite systems, discarding the
physical nature of the parts and of the environment, can exhibit
entangled stationary states towards the system can be guided by a
probabilistic scheme of measurement.
\section{Applications}
The best test of a new mathematical tool is its ability to
reproduce old results and to predict new results. In the contest
of old exemplary problem I have reproduced:

- the photocounting formula \cite{tesi,Davies}

- the enviroment-induced entanglement between two two-level
not-direct-interacting atoms placed in fixed arbitrary points in
the free space \cite{tesi,Plenio1999,Pellizzari,Ficek03}

- Carmichael unravelling of the master equation
\cite{tesi,Carmichael}\footnote{ Carmichael interpretation of
unravelling is made in a slight different way. He evidences the
possibility of different unravelling about which I'm trying to
better understand. Instead F.Petruccione and H.P.Breuer emphasize
the non selective-level  of measurement that imply an independent
-from a particular choice of the environment measured observable-,
unique unravelling. This way to think go to my same direction.}

Moreover I have tested the NuD theorem's predictive capability
solving the dynamics of:

- two two-level dipole-dipole interacting atoms placed in fixed
arbitrary points inside a single mode cavity in presence of atomic
spontaneous emission and cavity losses \cite{NicolosiPRL}.

- N two-level not-direct-interacting atoms placed in fixed
arbitrary points inside a single mode cavity in presence of atomic
spontaneous emission and cavity losses \cite{Nicolosi}.

- A bipartite hybrid model, known as Jaynes-Cummings model,
constituted by an atom and a single mode cavity linearly coupled
and spontaneously emitting in the same environment (work in
progress)

- Two harmonic oscillator linearly coupled and spontaneously
emitting in the same environment (work in progress)

\section{Was the dynamical behaviour of markovian quantum systems predictable?
 The {\it open system in an open-observed environment} model}
The physical basis to answer this question is provided by
continuous measurement theory by virtue of which we can regard the
Markovian Master Equation as the evolutionary equation describing
an open quantum system, whose environment is continuously
monitored. {\it Indirect quantum measurements} scheme (Braginsky
and Khalili, 1992, \cite{indqm}) seems to go to the right
direction because it theorizes the solution of dynamical problem,
here analyzed, as statistical mixture of independent events
referred to as {\it non-selective} measurement. An {\it indirect
quantum measurement} can be viewed as consisting of three
elements. The first element is the quantum system of interest and
it is called {\it quantum object}. The second element is the
so-called {\it quantum probe} on which the measurement is
performed. The third element of the scheme is a classical
apparatus by which a measurement on the quantum probe is
performed.

If we identify the quantum probe as the environment and replace
the third element with a {\it quantum measurement ideal device}
(characterized by its own Hilbert space) able to detect all the
environment elementary excitations coming from the system and to
make a click as evidence of the detection, it is possible to show
that the Markovian Master Equation is also obtainable from the
microscopic model describing this ideal full-quantum measurement
scheme. The principal effect of this {\it quantum measurement
ideal device} is to induce loosing of memory in the environment
respect to the interaction with the system: every time the system
give an excitation to the environment the device capture it. I
name it {\it Memory-Cleaner} (M-C).

Let us suppose that a $t=0$ the environment is in its ground state
(thermal state at the temperature $T=0$), the M-C is in its ground
state \footnote{ The characterization of the spectrum of the M-C
is not relevant in this contest. One may  thinks of a
photoemissive source coupled to an environment interacting with
one or more than one ideal photon detector
\cite{F.Petruccione,Carmichael}.} and the system is in an
arbitrary excited state and let us ask: how does they evolve? Two
composite events may happen:

- the system does not give excitations to the environment; the M-C
doesn't capture excitations;

- the system looses one or more than one excitation in the
environment; the M-C captures excitations.

In the first case the M-C will not make a click, the environment,
the system and the M-C resulting undisturbed by their mutual
interactions. In the second case the M-C will produce one or more
than one click as evidence of the fact that the system has
exchanged energy with the environment and the environment has
interacted with the M-C. In this second case the state of the
system has changed because of the interaction with the
environment, the state of the M-C has changed because of the
interaction with the environment, instead the status of
environment has been reset, respect to the coupling to the system,
to its initial state because of the presence of M-C device. The
time involved in the reset process can be considered of the same
order of the coherence time characterizing a markovian
environment.

This qualitative analysis shows that the introduction of ideal M-C
realizes a thermal equilibrium condition maintaining the
environment temperature constant and equal to zero (deviations
from equilibrium can happen in a time scale so short respect to
the time in which the system change appreciably its state that we
can assume to look at the system evolution in a coarse-grained
time scale assuming the environment evolution to happen over time
which are not resolved).

From a microscopic point of view this situation is represented by
the hamiltonian
\begin{eqnarray}\label{M-C}
H=H_{M-C}+H_S+H_B+H_I+H_{B,M-C}
\end{eqnarray}
containing the free energy of the quantistic M-C ($H_{M-C}$), of
the system ($H_S$) and of the environment ($H_B$) and the
interaction hamiltonians ($H_I$ and $H_{B,M-C}$) describing
respectively the coupling between the M-C and the environment and
the coupling between the system and the environment. The reduced
density matrix of the system is obtainable by tracing over the
environment and M-C degrees of freedom the corresponding Liouville
equation
\begin{eqnarray}
\dot{\rho}(t)=-\frac{i}{\hbar}[H,\rho (t)]
\end{eqnarray}
where $\rho (t)$ is the total density matrix belonging to the
Hilbert space given by the tensor product of the Hilbert space of
the system, of the environment and of the M-C.

The thermal equilibrium condition in the environment, realized by
the M-C, ensure that
\begin{eqnarray}
\rho(t)=\rho_B(0)\rho_{S,M-C}(t)
\end{eqnarray}
where $\rho_B(0)$ is the initial Gibbs state ($T=0$) of the
environment and $\rho_{S,M-C}$ is the density matrix describing
the quantum evolution in the tensor product Hilbert space of the
system and the M-C. The trace over the M-C degrees of freedom
($Tr_{MC}\{\cdot\}$) give

\begin{eqnarray}
\rho_B(0)\dot{\rho_{S}}(t)=-\frac{i}{\hbar}[H_S,\rho_B(0)\rho_{S}(t)]
-\frac{i}{\hbar}[H_B,\rho_B(0)\rho_{S}(t)]\\\nonumber
-\frac{i}{\hbar}[H_I,\rho_B(0)\rho_{S}(t)]-\frac{i}{\hbar}Tr_{MC}\{[H_{B,M-C},\rho(t)]\}
\end{eqnarray}
and then the trace over environment degrees of freedom ($Tr_B \{
\cdot \}$) give
\begin{eqnarray}
\dot{\rho_{S}}(t)=-\frac{i}{\hbar}[H_S,\rho_{S}(t)]
-\frac{i}{\hbar}Tr_B\{[H_I,\rho_B(0)\rho_{S}(t)]\}\\\nonumber
-\frac{i}{\hbar}Tr_B\{Tr_{MC}\{[H_{B,M-C},\rho(t)]\}\}.
\end{eqnarray}
The last term in the above equation is zero. This equation has to
describes a quantum system S coupled to a memory-cleaned
environment B. The environment characterization ensures that
$\rho_S(t)$ satisfies the eq. (\ref{ME0}) known as Markovian
Master Equation. Actually, this equation takes into account both,
the quantum M-C device and the environment through the Markov and
the Born approximations that can be seen as the memory-cleaned
environment approximation and the environment equilibrium state
assumption. Moreover, the introduction of the ideal M-C, beside to
give a full justification of Born-Markov approximation, induces to
think that an {\it almost} memory-cleaned environment is modeled
by a not-ideal M-C which limit case, a full memory environment, is
modeled by a not-working M-C, a device unable to capture
excitations (to interact with environment) or, equivalently, by a
closed microscopic model containing only the system, the
environment and their interaction. In this way exact open system
dynamics can be obtained by eq. (\ref{M-C}) just removing the M-C
and its interaction. On the ground of this considerations it is
possible to divide the eq. (\ref{ME0}) in three terms:

- we can recognize in the {\it sandwich} terms
\begin{eqnarray}
\sum_{\omega
>0 ,\alpha ,\beta} \gamma^{\alpha ,\beta} (\omega)A^{\beta}
(\omega) \rho_i (t)A^{\alpha \dag} (\omega)
\end{eqnarray}
the jump terms due to the M-C device and describing the capture of
an excitation of frequency $\omega$ and the consequent projection
(jump) of the vector state into the subspace characterized by a
free energy lower of $\omega$ than the previous of the measurement
act.

-  we can recognize in the {\it free energy constant} terms
\begin{eqnarray}
-\frac{1}{2}\sum_{\omega
>0 ,\alpha ,\beta} \gamma^{\alpha ,\beta} \{A^{\beta \dag} (\omega) A^{\alpha}
(\omega),\rho_i(t) \})
\end{eqnarray}
the effective dissipative part responsible of going to zero of
populations and coherences of the excited states of the open
system of interest and due to the {\it open observed environment}
we are considering. In fact these terms introduce an imaginary
frequency in the free-energy spectrum of $H_S$, so generating an
exponential decaying of excited free eigenstates.

-  we can recognize in the {\it free energy Lamb-Shift} terms
\begin{eqnarray}
\sum_{\omega
 ,\alpha ,\beta} S^{\alpha ,\beta} [A^{\beta \dag} (\omega) A^{\alpha}
(\omega),\rho_i(t) ]
\end{eqnarray}
the environment-induced multipartite cooperation not-vanished by
the loss of memory of the environment.

The found solution (NuD theorem) tell us that the state of the
system is a statistical mixture of the free energy system
eigenoperators. This fact depends and it is consistent with the
existence of the M-C measurement device because the act of
measurement introduces a stochastic variable respect to which we
can only predict the probability to have one or another of the
possible alternative measures. These probabilities can be regarded
as the weight of the possible alternative generalized trajectories
and, analytically, they are given by the partial traces of the
$\rho_i$. With this approach the dynamics has to be depicted as a
statistical mixture of this alternative generalized trajectories.

Moreover the found trajectories evolve in time in a deterministic
way: for example the trajectory relative to the initially excited
system state is a shifted free evolution characterized by complex
frequencies that means an exponential decaying free evolution.
This statement may give the sensation that every system has to
decay in its ground state because of the observed dynamics. It is
in general not true. Actually, if the system is multipartite, it
is possible that it admits excited and entangled equilibrium
Decoherence Free Subspace (DFS) (so as it happens in a lot of
known models), constituted by states on which the action of $H_I$
is identically zero and then, if the system, during evolution,
passes through one of these states, the successive dynamics will
be decoupled from the environment evolution. An equilibrium
condition is reached in which entanglement is embedded in the
system \cite{Nicolosi,NicolosiPRL}.

What could ensure, for example, that an entangled decoherence-free
state, if existing, has been generated? (For example) the number
of click we hear in a period of time long enough respect
spontaneous emission rate. Actually, if the numbers of the clicks
is less than the numbers of initial excitations then we can say
that it has been generated a decoherence-free state
\cite{Nicolosi,NicolosiPRL,Plenio1999}.

If, on the contrary, the system is monopartite it is possible to
demonstrate that the only possible DFS is generated by the ground
state of the system so that a monopartite system will loose its
internal coherence and the population of excited states because of
the measurement and the consequent interaction with a
memory-cleaned environment, unable to induce cooperation among the
parts: there exists only one part.

In this case the numbers of clicks has to be the same of initial
system excitations. The M-C {\it sounds} as a quantum logic
counter.


\begin{thebibliography}{99}
%%%Quantum register porte logiche

\bibitem{F.Petruccione} F.Petruccione, H.P.Breuer, {\sl The Theory of open quantum
sistem},
Oxford University Press (2002)

\bibitem{Gardiner} C.W.Gardiner, {\sl Quantum Noise}, Springer-Verlag Berlin
(1991)

\bibitem{Louiselle}W.H.Louiselle, {\sl
Quantum statistical properties of radiation}, Ed. John Wiley e
Sons (1973)

\bibitem{Alicki}R.Alicki, K.Lendi, {\sl Quantum Dynamical Semigroups and Applications}, Lectures Notes in Physics {\bf V286},
(Springer-Verlag 1987)

\bibitem{Carmichael} H.Carmichael, {\sl An Open System Approach to Quantum Optics}, Lectures Notes in Physics {\bf m18},
(Springer-Verlag 1993)

\bibitem{Molmer1}Y.Castin, K.Molmer, {\sl Phys. Rev. Lett.} {\bf 74},
3772 (1995)

\bibitem{Molmer2}J.Dalibard, Y.Castin, K.Molmer, {\sl Phys. Rev. Lett.} {\bf 68},
580 (1992)

\bibitem{Davies} E.B.Davies, {\sl Commun. Math. Phys.} {\bf 15},
277 (1969)

\bibitem{Davies2} M.D.Srinivas, E.B.Davies, {\sl Opt. Acta} {\bf 28},
981 (1981)

\bibitem{Plenio-Knight} M.B.Plenio, P.L.Knight, {\sl Rev. Mod. Phys.} {\bf 70},
101 (1998)

\bibitem{diagramma} F.Petruccione, H.P.Breuer, {\sl Concepts and methods in the theory of open quantum systems systems},
 Lectures Notes in Physics {\bf V622},  Springer-Verlag Berlin (2003)


\bibitem{Carm2000}H.J. Carmichael and K. Kim,
{\sl Optics Commun.} {\bf 179}, 417 (2000)


\bibitem{Carm2003}J.P. Clemens, L. Horvath, B.C. Sanders, and H.J. Carmichael,
{\sl Phys. Rev. A} {\bf 68}, 023809 (2003)


\bibitem{indqm} V.B.Braginsky, F.Ya.Khalili, {\sl Quantum
Measurement},
Cambridge University Press (1992)

\bibitem{Nicolosi} S. Nicolosi, A. Napoli, A. Messina,   {\sl quant-ph0306104}
(2003), submitted to EPJD

\bibitem{NicolosiPRL} S. Nicolosi, A. Napoli, A. Messina , F.Petruccione  {\sl quant-ph0402211}
(2004), axcepted on PRA

\bibitem{Barnett2002}S.M.Barnett, P.M.Radmore, {\sl
Methods in Theoretical Quantum Optics}, Oxford Series in Optical
and Imaging Science, {\bf 15} (2002)

\bibitem{Leonardi}C.Leonardi, F.Persico, G.Vetri, {\sl La rivista del Nuovo Cimento}{\bf
9}(1986) 3,4

\bibitem{subradiant state}G.Benivegna A.messina, {\sl J.Mod.Opt.} {\bf 36} (1989)
1205

\bibitem{Kozierowski}M.Kozierowski, S.M.Chumakov, A.A.Mamedov, {\sl Journal of Mod. Opt.} {\bf 40}
(1993)453

\bibitem{Shore}B.W.Shore, P.L.Knight, {\sl Journal of Mod. Opt.} {\bf 40}
(1993)7

\bibitem{Kudryavtsev}I.K.Kudryavstev, A.Lambrecht, H.Moya-Cessa, P.L.Knight, {\sl Journal of Mod. Opt.} {\bf 40}
(1993)8

\bibitem{Phoenix}S.J.D.Phoenix, S.M.Barnett, {\sl Journal of Mod. Opt.} {\bf 40}
(1993)6

\bibitem{Dicke}Dicke, {\sl phys. rep.} {\bf 93} (1954)99

\bibitem{Nielsen} M. A. Nielsen and I. L. Chuang, \textit{Quantum
    Computation and Quantum Information}, Cambridge University Press,
  Cambridge, (2000).

\bibitem{Cinesi} Guo Ping Guo et al, {\sl Phys. Rev. A}
{\bf 65}, 042102 (2002)

\bibitem{Cinesi2} Lu-Ming Duan et al, {\sl Phys. Rev. A}
{\bf 58},       (1998)

\bibitem{Cirac} J.I.Cirac, P.Zoller, {\sl Phys. Rev. A}
{\bf 50}, R2799 (1993)

\bibitem{Shneider} Shneider, Milburn, {\sl Phys. Rev. A}
{\bf 65}, 042107 (2002)

\bibitem{Yang} G.J.Yang, O.Zobay, P.Meystre, {\sl Phys. Rev. A}
{\bf 59}, 4012 (1998)

\bibitem{Hagley} E.Hagley, X.Maitre, G.Nogues,
C.Wunderlich, M.Brune, J.M.Raimond, S.Haroche, {\sl Phys. Rev.
Lett.} {\bf 79}, 1 (1997)

\bibitem{Foldi} P.Foldi, M.G.Benedict, A.Czirjak, {\sl Phys. Rev. A}
{\bf 65}, 021802 (2002)

\bibitem{tesi} S.Nicolosi, {\sl Soluzione operatoriale esatta di master equation ottiche generalizzate: applicazioni a problemi di interazione radiazione materia},
Degree Thesis (2003), unpublished

\bibitem{Plenio1999} Plenio, Huelga, Beige, Knight, {\sl Phys. Rev. A} {\bf 59} (1999)
2468

\bibitem{Pellizzari} T. Pellizzari, S.A. Gardiner, J.I. Cirac,
P.Zoller, {\sl Phys. Rev. Lett.} {\bf 75}, 3788 (1995)

\bibitem{Ficek03} Z.Ficek, R.Tanas, {\sl quant-ph 0302124}



\end{thebibliography}
\end{document}